\documentclass[letterpaper, 10 pt, conference]{ieeeconf}  % Comment this line out if you need a4paper

\IEEEoverridecommandlockouts                              % This command is only needed if 
\makeatletter
\def\endthebibliography{%
	\def\@noitemerr{\@latex@warning{Empty `thebibliography' environment}}%
	\endlist
}
\makeatother

\usepackage{graphics} % for pdf, bitmapped graphics files
\usepackage{epsfig} % for postscript graphics files
\usepackage{mathptmx} % assumes new font selection scheme installed
\usepackage{times} % assumes new font selection scheme installed
\usepackage{amsmath} % assumes amsmath package installed
\usepackage{amssymb}  % assumes amsmath package installed
\usepackage{siunitx}
\usepackage{verbatim}
\usepackage{mathrsfs}
\usepackage{booktabs}
\usepackage{xcolor}
\usepackage{flushend}
\usepackage{url}
\usepackage[ruled,vlined]{algorithm2e} % package for algorithm
\title{\LARGE \bf  Application of Ghost-DeblurGAN to Fiducial Marker Detection}

\author{Yibo Liu, Amaldev Haridevan, Hunter Schofield, Jinjun Shan% <-this % stops a space
\thanks{The authors are with Department of Earth and Space Science and Engineering, York University, Toronto, Ontario M3J1P3, Canada 
        {\tt\footnotesize \{yorklyb,amaldev,hunterls,jjshan\}@yorku.ca}}%
}

\begin{document}

\maketitle
\thispagestyle{empty}
\pagestyle{empty}

%%%%%%%%%%%%%%%%%%%%%%%%%%%%%%%%%%%%%%%%%%%%%%%%%%%%%%%%%%%%%%%%%%%%%%%%%%%%%%%%
\begin{abstract}
Feature extraction or localization based on the fiducial marker could fail due to motion blur in real-world robotic applications. To solve this problem, a lightweight generative  adversarial  network, named Ghost-DeblurGAN, for real-time motion deblurring is developed in this paper. Furthermore, on account that there is no existing deblurring benchmark for such task, a new large-scale dataset, YorkTag, is proposed that provides pairs of sharp/blurred images containing fiducial markers. With the proposed model trained and tested on YorkTag, it is demonstrated that when applied along with fiducial marker systems to motion-blurred images, Ghost-DeblurGAN improves the marker detection significantly. The datasets and codes used in this paper are available at: https://github.com/York-SDCNLab/Ghost-DeblurGAN.

\end{abstract}

% \fi
\IEEEpeerreviewmaketitle

\section{Introduction}
\begin{figure}[htpb]
	\centering
	\includegraphics[width=3.35in]{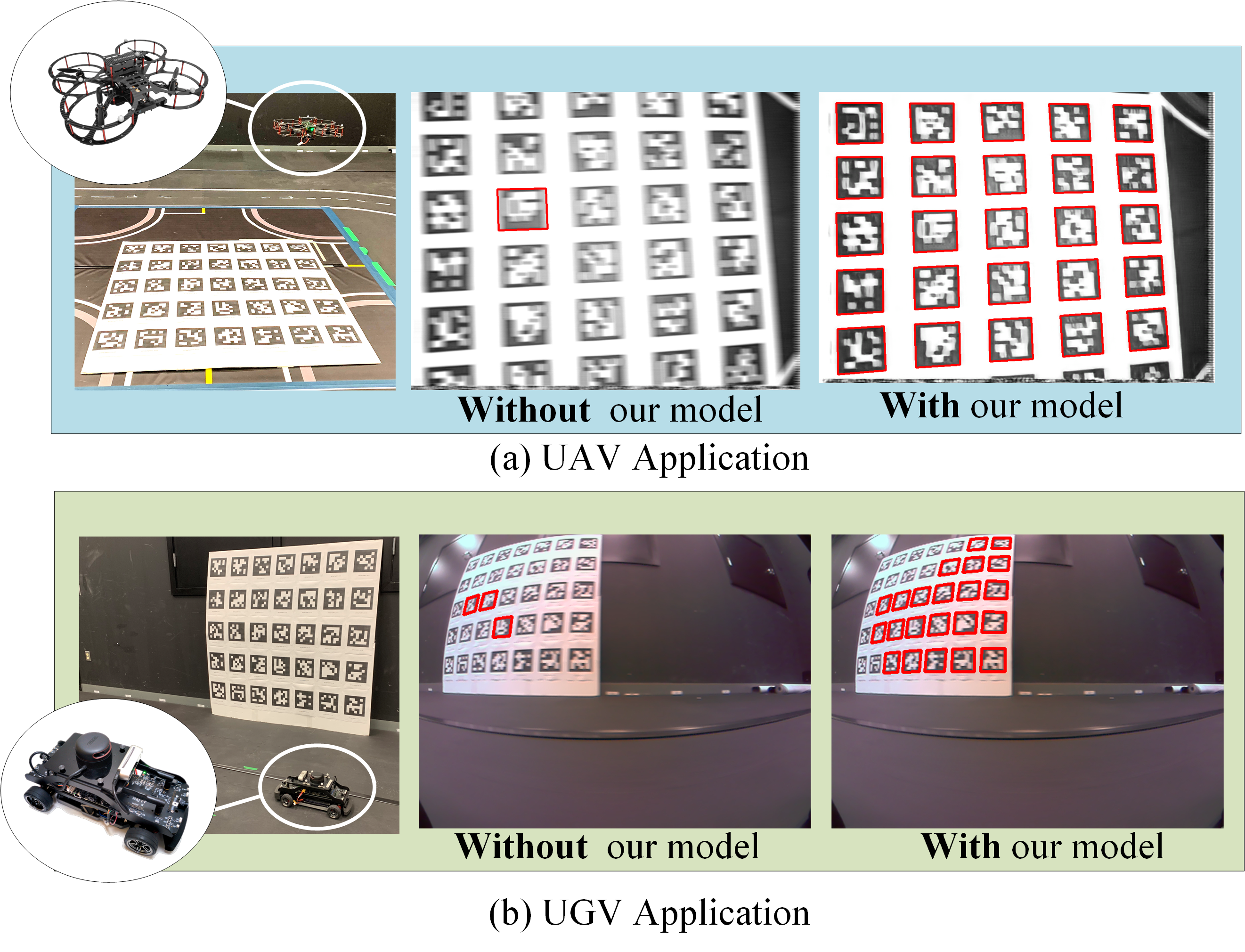}
%	\vspace{-0.1in}
	\caption{Visual comparison of marker detection with and without Ghost-DeblurGAN in robotic applications. Detected markers are labeled by red frames. (a):  A frame from a video captured by a downwards camera onboard a maneuvering UAV (Qdrone, from Quanser Inc.). (b): A frame from a video captured by a low-cost CSI camera onboard a moving UGV (Qcar, from Quanser Inc.). }
	\label{fig1}
\end{figure}

Motion blur resulting from camera shake and rapid object motion is frequently seen in photos or videos captured by, for instance, hand-held cameras as well as cameras onboard autonomous vehicles \cite{deblurgan1,deblurgan2,nah}. Severe motion blur causes image degradation, which hinders the object detection task and could lead to detection failure \cite{bert}. In recent years, deep-learning-based deblurring methods \cite{deblurgan1,nah,sayed} have been proven effective in improving object detection performance when used along with object detection systems, such as YOLO \cite{redmon}.  However, previous studies have not dealt with the application of deep-learning-based deblurring methods in fiducial marker systems. Fiducial markers \cite{wang,ap3,aruco} are particular objects, commonly planar and  black-and-white, that provide the environment with controllable artificial features. They are widely used in augmented reality and robotic systems \cite{olson,ap3}. Although the theories behind marker detection \cite{wang,ap3,aruco} and the deep-learning-based object detection \cite{redmon,bert} are different, as shown in Fig.~\ref{fig1}, marker detection also suffers from motion blur.

Marker detection is the first step and a fundamental requirement of high-level tasks \cite{munoz,munoz2019,yibo}. Suppose that the vehicles in Fig.~\ref{fig1} utilize fiducial markers to acquire artificial features for marker-based Simultaneous Localization and Mapping (SLAM) \cite{munoz,munoz2019}, then the feature extraction will fail since the markers cannot be detected due to motion blur. Similarly, suppose that the vehicles in Fig.~\ref{fig1} employ the markers as 3D fiducials to obtain localization information directly for a visual servoing framework \cite{yibo}, the vehicle will lose its signal input (localization information) when the image is blurred. To the best of our knowledge, CCTag \cite{cctag} is the only fiducial marker system that considers motion blur in its algorithm, however, the other visual fiducial marker systems \cite{olson,wang,aruco,ap3} do not take motion blur as a routine case, which makes the adoption of the deblurring method vital in real-world applications involving fiducial markers.

Despite deblurring performance rankings on public datasets, including GoPro \cite{nah}, being updated nearly on a monthly basis, the model sizes and runtimes of the state-of-the-art (SOTA) approaches \cite{chen2021hinet,zamir} tend to increase as well. This is unfriendly to real-time applications. On the contrary, the SOTA fiducial marker systems, such as AprilTag 3 \cite{ap3}, have already achieved real-time detection. Hence, the notable gap between deep-learning-based deblurring methods and fiducial marker systems is the runtime performance. To fill this gap, factors, including model size, FLOPs, and inference time, rather than solely the SOTA deblurring quality, need to be considered.

This work aims to solve the above-mentioned problems and the contributions of this work are summarized as follows. 1) The development of a novel lightweight generative adversarial network, Ghost-DeblurGAN, for real-time deblurring. Although the proposed model is not the SOTA deblurring method compared with other non-lightweight implementations \cite{chen2021hinet,zamir}, Ghost-DeblurGAN is more appropriate for real-time applications as it has a tiny model size and rapid inference time. 2) The proposal of a new large-scale dataset, \textbf{YorkTag}, that contains blurred images containing fiducial markers and their corresponding ground truth sharp images. The \textbf{YorkTag} can be used as a new benchmark for future deblurring research.

\section{Related Work\label{relate}}

\subsection{GAN-Based Deblurring}
A generative adversarial network (GAN) is an unsupervised deep-learning framework proposed by Goodfellow et al  \cite{gan}. Ramakrishnan et al. \cite{deep} first introduce a GAN to deblurring tasks, which indicates recovering a sharp image from a blurred image. Kupyn et al. proposed DeblurGAN-v1 in \cite{deblurgan1} and DeblurGAN-v2 in \cite{deblurgan2}. To avoid the possible gradient explosion, DeblurGAN employs Wasserstein GAN \cite{wgan} and DeblurGAN-v2 adopts modified Least Squares GAN  \cite{lsgan}. Zhang et al. \cite{dbgan} proposed learning-to-DeBlur GAN (DBGAN) \cite{dbgan} which has a better deblurring quality than DeblurGAN-v2 on the GoPro dataset. However, DBGAN \cite{dbgan} is developed neither as a lightweight implementation nor for real-time deblurring purposes while DeblurGAN-v2 (MobileNetV2) \cite{deblurgan2} achieves real-time deblurring with a tiny model size. As a matter of fact, DeblurGAN-v2 is not the SOTA deblurring method and GAN is definitely not the only choice for deblurring, for instance, HINet \cite{chen2021hinet} and MPRNet \cite{zamir} already exceed it in terms of accuracy. Nevertheless, the DeblurGAN-v2 (MobileNetV2) still belongs to SOTA lightweight deblurring networks for real-time applications.

There is some incremental work \cite{truong2020slimdeblurgan,deblurgan2} for the sake of better real-time deblurring based on the DeblurGAN \cite{deblurgan1,deblurgan2}. A slimmed version of DeblurGAN-v1, called SlimDeblurGAN is introduced in \cite{truong2020slimdeblurgan} by adopting the network slimming approach introduced in \cite{liuslim} on DeblurGAN-v1 \cite{deblurgan1}. However, DeblurGAN-v1 no longer belongs to the SOTA deblurring methods while SlimDeblurGAN shows accuracy degradation compared to it \cite{truong2020slimdeblurgan}. In \cite{deblurgan2}, Kupyn et al. propose a radical attempt named  DeblurGAN-v2 (MobileNet-DSC) where all the normal convolutions are replaced by depthwise convolutions \cite{dw} in DeblurGAN-v2 (MobileNetV2). The aim of this approach is to further boost the deblurring efficiency, which is similar to the objective this work. Compared with DeblurGAN-v2 (MobileNetV2), the efficiency of MobileNet-DSC is promoted, but again at the cost of deblurring quality degradation \cite{deblurgan2}.

\subsection{Marker Detection}
The typical marker detection process includes image binarization, boundary segmentation, candidate quad detection, and decoding \cite{ap3,aruco}. The motion blur can impede marker detection in two ways. First, given that a marker needs to be detected as a candidate quad before finally being determined as a valid marker \cite{ap3}, marker detection will fail before the decoding step due to a quad detection failure if the line and quad features are ruined by severe motion blur. Secondly, even if a marker is detected as a candidate quad, the decoder will filter the marker out if its binary payload is found in the known code words \cite{ap3}, while the payload can be wrecked by motion blur. The improvement of YOLO-based object detection brought by DeblurGAN-v1 shown in \cite{deblurgan1} indicates the potential of deblurring approaches in improving marker detection.

\section{Network Architecture}
\begin{figure*}
\centering
\includegraphics[width=16cm]{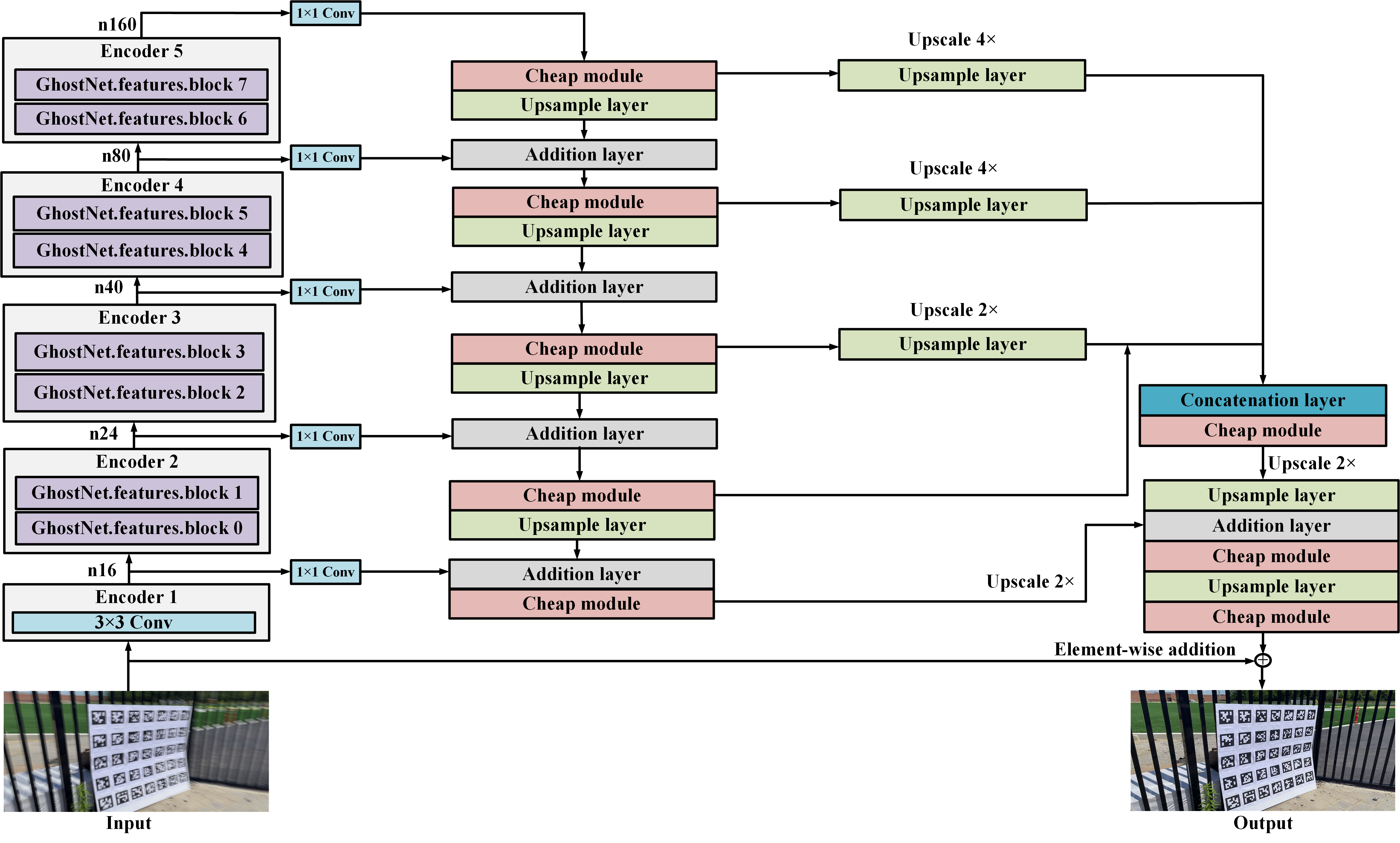}
%\vspace{-0.1in}
\caption{Architecture of the Ghost-DeblurGAN generator.}
\label{overall}
\end{figure*}

The following sections illustrate the course of designing the generator. After four trial models, as shown in Section \ref{exp}, the optimal one that outperforms the DeblurGAN-v2 (MoibleNeV2) \cite{deblurgan2} in all terms is proposed.

\subsection{Generator Design}
DeblurGAN-v2 adopts the FPN \cite{fpn} architecture to build the generator. If we utilize more complex architecture upgraded from FPN, such as Feature Pyramid Grids \cite{fpg}, the deblurring quality should be improved, as it can provide features with richer semantic and quality information \cite{fpg}. However, at the same time, the inference time of the network will also increase due to the more complex inner connections or more pathways, which is contrary to our intention of improving the real-time deblurring efficiency. Hence, the FPN architecture is also adopted in this work. For the FPN architecture, the design starts from the backbone. In recent years, significant progress has been made in the backbone feature extractor. Quite a few novel lightweight backbones, such as MobileNetV3 \cite{mobilenetv3} and GhostNet \cite{ghostnet} are proposed since MobileNetV2 \cite{mobilenetv2}, which is the backbone adopted in the lightweight implementation of DeblurGAN-v2 \cite{deblurgan2}. 

The current SOTA backbones,  MobileNetV3 and GhostNet, employ the same architecture which is automatically searched by platform-aware NAS  \cite{mobilenetv3,ghostnet}. MobileNetV3 and GhostNet have neck and neck performance in ImageNet Classification and Object Detection, but it should be noted that MobileNetV3 and GhostNet do not show accuracy superiority over MobileNetV2 in Object Detection; however, their virtue is in the reduced FLOPs \cite{ghostnet}.

Yet, it is to be verified whether MobileNetV3 and GhostNet have better deblurring quality than MobileNetV2 in the architecture of DeblurGAN-v2. The comparison results are presented in Section~\ref{exp}. In DeblurGAN-v2, the last few layers, including average pooling and full connections, of the backbone are abandoned since these layers are originally developed for the image classification. However, we experimentally found that for the DeblurGAN-v2 architecture, the last eight layers of GhostNet can be abandoned and by doing so the accuracy is even boosted. Hence, as seen in Fig.~\ref{overall}, only the first eight feature blocks of GhostNet are preserved in the proposed model and the dimension of the output tensor of the last layer is 160 instead of 960, which is different from the original GhostNet. Moreover, the remaining top-down pathways and horizontal connections in the FPN are modified correspondingly afterward. For example, the upscale ratios of the proposed generator are 4,4,2,2,2 from top to bottom since the proposed lightweight model is shallow compared to the original model which adopts the upscale ratios 8,4,2,1,1. Refer to \cite{deblurgan2} to check the detailed difference between the proposed generator and the original one.

\subsection{Network Lightening and Accuracy Boost \label{light}}
It is guaranteed that if GhostNet or MobileNetV3 is adopted as the backbone, the FLOPs of the proposed model will be reduced compared with DeblurGAN-v2 (MobileNetV2). However, as demonstrated in Section~\ref{exp}, the FLOPs reduction is far from  satisfactory. In this section, the modifications in the top-down pathways are introduced, which significantly reduce the FLOPs, and improve performance. 
\begin{figure}[thb]
	\centering
	\includegraphics[width=3.3in]{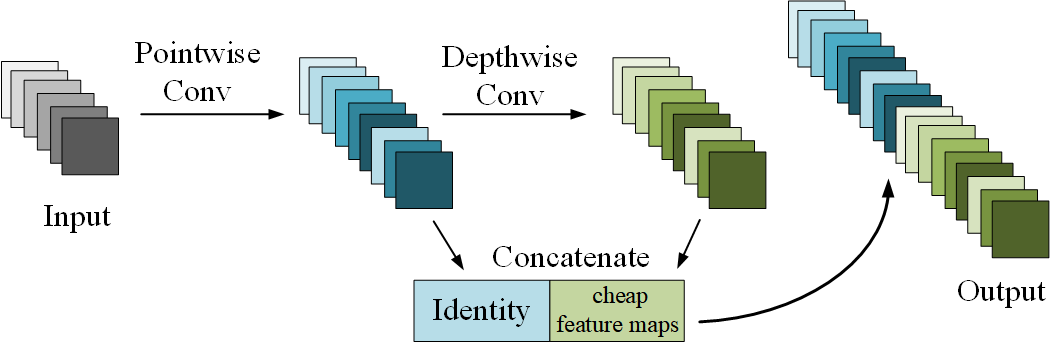}
    	\caption{Cheap module.}
	\label{GM}
\end{figure}

Inspired by \cite{ghostnet} and the incremental work (MobileNet-DSC) introduced in \cite{deblurgan2}, the cheap modules shown in Fig.~\ref{GM} are adopted instead of the conventional 2D convolutional layers in the top-down pathways. The cheap module first generates intrinsic feature maps using pointwise convolutions \cite{pw} and then adopts depthwise separable convolutions \cite{dw} to the intrinsic feature maps to generate cheap feature maps (denoted in green in Fig.\ref{GM}).  After that, the intrinsic and cheap feature maps are concatenated to be the output feature maps where half of them are cheap feature maps. As shown in Table~\ref{tab1}, the FLOPs of the model is reduced by \textbf{53.12\%} due to the lightening.
 \begin{figure}[thpb]
	\centering
	\includegraphics[width=3.3in]{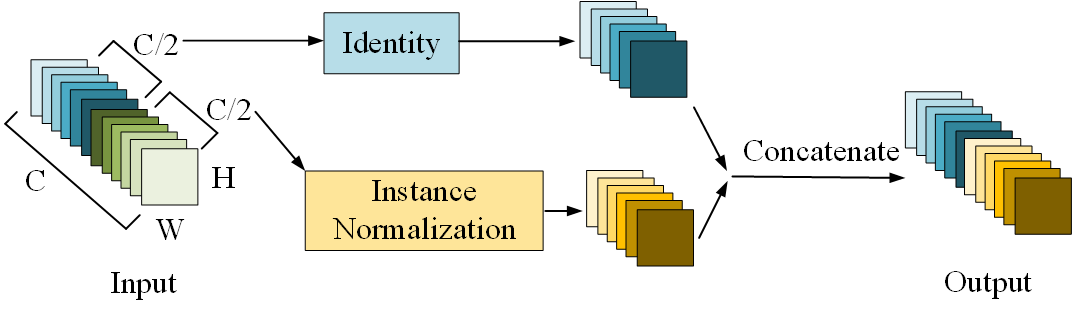}
	\caption{Half Instance Normalization.}
	\label{hin}
\end{figure}

However, as illustrated in \cite{deblurgan2}, simply replacing the conventional convolutions with depthwise separable convolutions will cause an accuracy degradation rather than an accuracy boost. It is experimentally found that this is the same case for the cheap module shown in Fig.~\ref{GM}. Hence, to further improve the accuracy, the following modification is conducted. Inspired by \cite{chen2021hinet}, the layers shown in Fig.~\ref{hin} are adopted following the cheap module.  In particular, the input feature tensor $F_{in}\in\mathbb{R}^{C_{in}\times H \times W}$ is first devided into two parts in the channel dimension: $F_{1}\in\mathbb{R}^{C_{in}/2\times H \times W}$ and $F_{2}\in\mathbb{R}^{C_{in}/2\times H \times W}$. The first part $F_{1}$ is inputted into the conventional Instance Normalization layer \cite{in} and then concatenates with $F_{2}$ to compose the output feature tensor.  The normalization layer shown in Fig.~\ref{hin} preserves more context information from the remaining half channels, which is friendly to features in shallow layers \cite{chen2021hinet}. As shown in Table~\ref{tab1}, the accuracy is boosted, which is opposite to conventional incremental work \cite{truong2020slimdeblurgan,chen2021hinet}. Thereafter, the double-scale discriminator and loss function introduced in \cite{deblurgan2} are adopted to compose the GAN. 

\begin{table*}[htbp]
\caption{performance and efficiency comparison on the GoPro test dataset}
\begin{center}
\begin{tabular}{c|c|c|c|c|c|c}
\hline\hline
%&\multicolumn{3}{|c|}{\textbf{Table Column Head}} \\
Framework & Backbone&Backbone FLOPs$^{\mathrm{a}}$ &Overall FLOPs$^{\mathrm{a}}$&Time on GPU &  PSNR &  SSIM   \\
\cline{1-7} 
DeblurGAN-v2 \cite{deblurgan2} & MobileNetV2 $1.0\times$ & 5.01G &43.75G& 60ms&  28.40 &  0.917 \\ \hline
Ghost-DeblurGAN-v1& MobileNetV3 $1.0\times$ & 4.31G &38.07G& 40ms& 28.05 & 0.918 \\ \hline
Ghost-DeblurGAN-v2&GhostNet  $1.0\times$ & \textbf{2.67G} &37.92G& 43ms& 28.42 &  0.913 \\ \hline
Ghost-DeblurGAN-v3&Modified GhostNet  $1.0\times$ & \textbf{1.94G} &37.76G& 42ms& 28.51 &  0.916 \\ \hline
Ghost-DeblurGAN-v4&Modified GhostNet  $1.0\times$& \textbf{1.94G} &\textbf{20.51G}& \textbf{37ms}&  \textbf{28.75} &  \textbf{0.919} \\ \hline\hline
\multicolumn{7}{l}{$^{\mathrm{a}}$ Note that the backbone and overall FLOPs are determined by the input tensor size. The values are with respect to the image size of the} \\
\multicolumn{7}{l}{GoPro dataset.}
\end{tabular}
\label{tab1}
\end{center}
\end{table*}
%\vspace{-0.2in}

\subsection{Training Dataset}
\begin{figure}[thpb]
	\centering
	\includegraphics[width=3.3in]{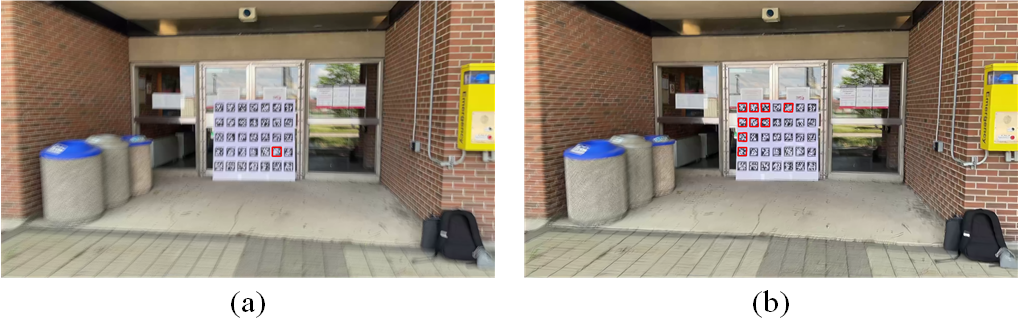}
%	\vspace{-0.15in}
	\caption{Images with the AprilTag 3 detector \cite{ap3}. a) blurred image. One marker is detected. b) Deblurred image from HINet \cite{chen2021hinet} (the SOTA deblurring approach) trained on the GoPro dataset. Six markers are detected.}
	\label{why}
\end{figure}

The GoPro dataset \cite{nah} is a common benchmark for deblurring. The sharp images are frames from videos captured by a GoPro Hero 4, and the blurred images are generated through averaging consecutive frames. The dataset is composed of 2103 pairs of sharp/blurred images for training and 1111 pairs for testing. Despite the different scenes involved, the network trained on a training dataset can achieve good deblurring quality when applied to a test dataset. However, the current deblurring benchmarks only contain routine scenes including pedestrians, cars, buildings, and human faces, etc. To illustrate the necessity of proposing a new deblurring benchmark containing fiducial markers, we test HINet \cite{chen2021hinet} which has the SOTA performance on the GoPro dataset with a blurred image and apply the Apriltag 3 detector \cite{ap3} to the deblurred image. As shown in Fig.~\ref{why}, due to the fact that HINet is trained on the GoPro dataset which contains no fiducial markers, the marker detection rate is far from  satisfactory.

\begin{figure}[thpb]
	\centering
	\includegraphics[width=3.3in]{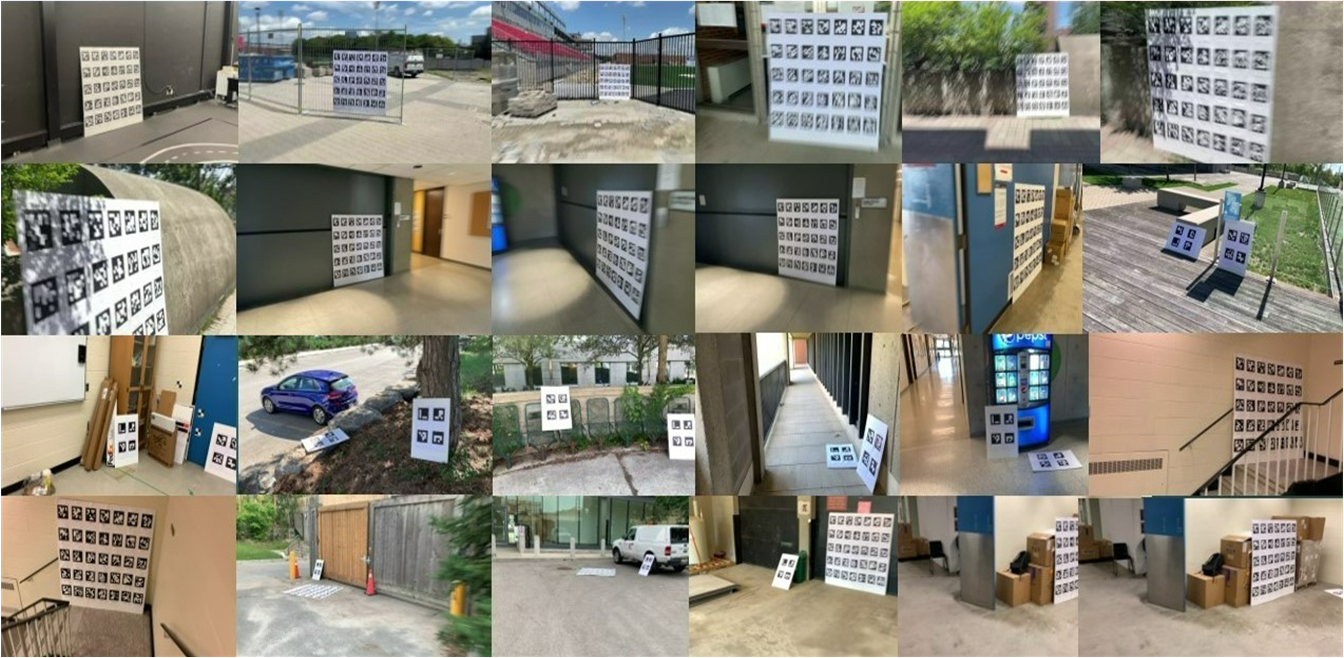}
	\caption{Some scenes of the proposed YorkTag dataset.}
	\label{yorktag}
\end{figure}

To end this, we propose a new large-scale dataset, \textbf{YorkTag}, that provides paired blurred and sharp images containing AprilTags \cite{ap3} and ArUcos \cite{aruco}. The proposed YorkTag can also be utilized as a benchmark for general deblurring research. As seen in Fig.~\ref{yorktag}, the indoor and outdoor scenes also contain objects including shelves, trees, cars, and so on. Moreover, the scale of YorkTag is close to that of GoPro. In YorkTag, the training set consists of 1577 image pairs, and the test set consists of 497 image pairs totalling 2074 blurry-sharp image pairs. 

In particular, the sharp images are from the high-resolution videos captured by iPhone 12 with the DJI OM 4 stabilizer. The blurred and sharp image pairs are generated through a similar process as in \cite{nah}. As the camera sensor receives light during a long exposure, an image over the time interval is accumulated, producing the blurry image \cite{blurgen}. This blurred image $B$ is formulated as follows:\begin{equation} 
	B=g\left(\dfrac{1}{T}\int^{T}_{t=0} S(t) dt\right) \simeq g\left(\dfrac{1}{N}\sum^{N-1}_{i=0} S[i]\right) 
\end{equation}
where $T$ and $S(t)$  denote the exposure time and the sensor signal of an image at time $t$, respectively. $N$ and $S[i]$ are the number of frames, and the $i^{th}$ frame signal captured during the exposure time. $g$ is the Camera Response Function  (CRF) that maps a signal $S(t)$ into an observed image $I(t)$ such that $I(t) = g(S(t))$. When the CRF is unknown we can approximate CRF as a gamma curve with $\gamma$= 2.2 \cite{CRF}.\begin{equation} g(x)= x^{\frac{1}{\gamma}}\end{equation}

With the CRF given we produce the signals $S[i]$ using $g^{-1}\left(I[i]\right)$, where $I[i]$ is the $i^{th}$ image in the batch. The blurred image is produced using $(2)$ and the corresponding ground truth is selected as the middle frame of the batch. A varying batch size of images ranging from 3 to 11 frames is used to produce YorkTag.

\section{EXPERIMENTAL RESULTS}\label{exp}

The implementation of our model is based on PyTorch. All the training and experiments are conducted on a desktop with Intel Xeon W-1290P CPU and NVIDIA Quadro RTX 4000 GPU.

\subsection{Qualitative Evaluation}
To illustrate the functionality of the proposed Ghost-DeblurGAN, visual comparison of the marker detection with and without our model is conducted and the experimental video is available at: https://github.com/York-SDCNLab/Ghost-DeblurGAN. As shown in this video, marker detection is significantly improved in the deblurred images. Hence, the proposed model can resolve the problems of feature extraction failure or localization loss due to motion blur.

\subsection{Comparison on the GoPro Dataset}
The proposed Ghost-DeblurGAN is compared with the original lightweight implementations of DeblurGAN-v2 (MobileNetV2) \cite{deblurgan2} on the \textit{linear} GoPro test dataset \cite{deblurgan2,nah} in terms of peak signal-to-noise ratio (PSNR), Structural Similarity Index (SSIM), inference time, and FLOPs. For a fair comparison, every model is trained from scratch on the \textit{linear} GoPro training dataset \cite{deblurgan2,nah} for 2000 epochs with 2000 images for each epoch. Table \ref{tab1} presents the comparison results of the original DeblurGAN-v2 (MobileNetV2) \cite{deblurgan2} and the four trial models of Ghost-DeblurGAN.

Ghost-DeblurGAN-v1 and Ghost-DeblurGAN-v2 denote that the architecture illustrated in Fig.~\ref{overall} is adopted while the backbones are MobileNetV3 and GhostNet, respectively.  Despite the architectures of Ghost-DeblurGAN and DeblurGAN-v2 being different, both of them employ FPN. It is shown from the comparison between Ghost-DeblurGAN-v1,v2, and original DeblurGAN-v2 (MobileNetV2) that MobileNetV3 and GhostNet do not have evident superiority over MobileNetV2 in terms of deblurring quality and overall FLOPs. Nevertheless, the inference time is markedly reduced when MobileNetV3 or GhostNet is adopted. Moreover, degradation occurs in terms of PSNR when using MobileNetV3 as the backbone. Ghost-DeblurGAN-v3 represents the case where the original GhostNet backbone of  Ghost-DeblurGAN-v2 is replaced by the modified GhostNet backbone. Here, the modified  GhostNet backbone indicates that only the feature blocks 0-7 are used. Compared with Ghost-DeblurGAN-v2, where feature blocks 0-9 are used in the backbone, the performance of Ghost-DeblurGAN-v3 is upgraded in all metrics. Ghost-DeblurGAN-v4 implies that the structures for network lightening and accuracy boost introduced in Section \ref{light} are utilized in Ghost-DeblurGAN-v3. Dissimilar with MobileNet-DSC \cite{deblurgan2} and SlimDeblurGAN \cite{truong2020slimdeblurgan}, the proposed model (Ghost-DeblurGAN-v4) simultaneously reduces the FLOPs and improves the deblurring quality compared with DeblurGAN-v2 (MobileNetV2). In particular, the overall FLOPs of Ghost-DeblurGAN is lower than DeblurGAN-v2 (MobileNetV2) by \textbf{53.12\%}. Nevertheless, the PSNR of Ghost-DeblurGAN increases to 28.75 from 28.40 and SSIM to 0.919 from 0.917. The comparison of Ghost-DeblurGAN and the SOTA deblurring method, HINet \cite{chen2021hinet}, is introduced in the next section.

%\subsection{Improvement in Marker Detection}
\subsection{Comparison on the YorkTag Dataset}
It should be noted that the intention of this work is to propose a lightweight deblurring network that can be used along with the fiducial marker system in real-time so that problems such as localization loss due to motion blur are resolved. Referring to \cite{chen2021hinet}, the deblurring quality of the proposed model is inferior to the SOTA deblurring method, HINet, when compared using the GoPro benchmark \cite{nah}. Despite this, considering the intention of this work, the proposed model is more appropriate than HINet \cite{chen2021hinet} in real-time applications on account of the following reasons. The model size of HINet is around 354 MB while that of the proposed model is solely \textbf{6.08 MB}. The inference time of HINet is approx. 602 ms on the tested GPU while that of Ghost-DeblurGAN is 37 ms. Furthermore, it is to be verified whether HINet has significant superiority over the proposed Ghost-DeblurGAN in marker detection since compared with PSNR and SSIM, the marker detection rate is more meaningful in real-world applications. To answer this question as well as to demonstrate that the proposed model significantly improves marker detection when the image is blurred, the following experiment is conducted.

The YorkTag test dataset is utilized as the benchmark. HINet and Ghost-DeblurGAN are trained on the training set of YorkTag from scratch (2000 images for each epoch and 2000 epochs in total). Then HINet and the proposed model are respectively applied to the blurred images of the test set to obtain the deblurred sets. Then, the AprilTag 3 detector \cite{ap3} and ArUco detector \cite{aruco} are adopted on the sharp, blurred, and deblurred images of the test dataset, respectively. Results are shown in Table \ref{tabdetection}.
\begin{table}[htb]
\caption{Comparison of the number of detected fiducial markers in different image sets}
\begin{center}
\begin{tabular}{c|c|c}
\hline\hline
%&\multicolumn{3}{|c|}{\textbf{Table Column Head}} \\
Images Set & Detected Markers & Detection Rate \\ \hline
Sharp &9761 & 100\% \\ \hline
Blurred & 3123& 31.99\%\\ \hline
Deblurred (HINet \cite{chen2021hinet}) & 6118& 62.68\% \\ \hline
Deblurred (Ghost-DeblurGAN) & 5769& 59.10\% \\ \hline\hline
\end{tabular}
\label{tabdetection}
\end{center}
\end{table}

The number of detected fiducial markers in the sharp image set can approximately correspond to the true number since sharp images are used as the ground truth. Compared with the sharp image set, solely 31.99\% (3123/9761) of the fiducial markers are detected in the blurred image set, which indicates the impediment to marker detection brought by motion blur. The detection rate increases to 62.68\% when using HINet \cite{chen2021hinet} and 59.10\% when using Ghost-DeblurGAN. Given that the marker detection rate of the proposed model is slightly inferior to HINet, Ghost-DeblurGAN is still a better choice in many real-time applications due to the factors mentioned previously, including the model size (354 MB vs. 6.08 MB) and inference time (602 ms vs. 37 ms).

\section{CONCLUSIONS}
In this paper, a novel lightweight deblurring network, Ghost-DeblurGAN, and its application to the fiducial marker detection are introduced. Moreover, a large-scale deblurring benchmark, YorkTag, that contains fiducial markers in the scenes is proposed. Different from DeblurGAN-v2 (MobileNetV2-DSC) \cite{deblurgan2} and SlimDeblurGAN \cite{truong2020slimdeblurgan} which reduce the FLOPs while comprising accuracy, the proposed model shows superiority in all considered metrics, including the size of FLOPs, deblurring quality, and efficiency, over DeblurGAN-v2 (MobileNetV2) \cite{deblurgan2}. Although the proposed model is slightly inferior to the SOTA deblurring method, HINet \cite{chen2021hinet}, in terms of marker detection rate on the YorkTag test set, the virtue of our model is the tiny model size and rapid inference time, which makes it a more suitable choice for real-time applications. When Ghost-DeblurGAN is adopted in applications involving fiducial markers, for instance, SLAM \cite{munoz,munoz2019} and visual servoing \cite{yibo}, problems of feature extraction failure or localization loss due to motion blur can be resolved.

%%%%%%%%%%%%%%%%%%%%%%%%%%%%%%%%%%%%%%%%%%%%%%%%%%%%%%%%%%%%%%%%%%%%%%%%%%%%%%%%
%\section*{APPENDIX}
\section*{ACKNOWLEDGMENT}
The authors would like to thank Liangyu Chen and Kai Han for constructive suggestions. This work was supported in part by NSERC Alliance Program under Grant ALLRP 555847-20, and in part by Mitacs Accelerate Program under Grant IT26108.
%The authors would like to thank Kai Han for insightful discussions. 



%\bibliographystyle{IEEEtran}

\bibliographystyle{IEEEtran} 
\bibliography{reference} 
\end{document}